\documentclass[%
aip,
 amsmath,amssymb,
reprint%
]{revtex4-1}

\usepackage{bbold}
\usepackage{graphicx}
\usepackage{dcolumn}
\usepackage{bm}
\usepackage{xcolor} 

\usepackage[utf8]{inputenc}
\usepackage[T1]{fontenc}
\usepackage{mathptmx}

\begin{document}

\title{Incommensurate Magnetic Order in Itinerant Systems from 1-D Correlated Topological Bulk States}
\date{\today}

\author{Jean-Guy Lussier}
\email[Email address:]{jlussier@kent.edu}
\affiliation{Department of Physics, Kent State University, Kent, OH 44242}

\begin{abstract}
The superposition of two rotated 2-D topological Dirac Hamiltonians with the same spin direction achieves separation of variables into two distinct directional 1-D topological Dirac Hamiltonians.  {\color{black} The 1-D solutions reproduce a spin density wave (SDW) type of magnetic ordering with a finite correlation length.} When the 1-D edge state solutions are combined to form a many-body bulk state, the incommensurate (IC) magnetic ordering reciprocal space wave vector observed in diffraction experiments is an average of crystalline lattice wave vectors in a moving center of momentum reference frame. These topological magnetic correlations determined {\color{black}by the length scales present in the underlying lattice} appear {\color{black}in Chromium and $\gamma$-Fe metal as well as} in f-electron systems with itinerant character, like Ce-compounds (CePtSn, CePdSn, CeRhIn$_5$, CeRhSn$_3$, CeNiAsO) but also in the High-Tc BSSCO and in the helical magnet system MnSi. The phenomenological algorithm for the calculation of the IC wave vector satellites is given in details.
\end{abstract}
\maketitle{


The standard theory for the description of IC magnetic order in itinerant and metallic systems is the onset of SDW as a result of instabilities of an electron gas {\color{black}in its paramagnetic state} due to the presence of a Fermi surface\cite{overhauserPhysRev.128.1437}.  The IC magnetic wave vector, which for a sinusoidal modulation appears as a chiral pair, equals $2 k_f$ {\color{black}in an effect called nesting of the Fermi surface}. More recently field theory approach describes the evolution of the magnetic order of itinerant and metallic ferromagnets at low temperatures into antiferromagnetic/IC magnetic order\cite{brando2016metallic}. In such calculations, which requires the setting of an appropriate many-body action, the spin is introduced as an extra bosonic variable. Alternatively, the Dirac equation offers a theoretical framework where momentum and spin are treated simultaneously. Applications of a modified Dirac Hamiltonian to condensed matter has applications with topological insulators and may hint at its many-body aspect. In this paper, we explore the 2-D solutions of the Dirac Equation that suggest the existence of topological effects in itinerant and metallic systems displaying IC magnetic order.

The modified Dirac equation, as is presented by Shen\cite{shen2012topological}, has a $\boldsymbol{p}^2$-correction to the mass term controlled by a gap parameter $B$ shown in Eq.\ref{topoDirac}.

\vspace{-0.0cm}
\begin{eqnarray} \label{topoDirac}
H= v \; \boldsymbol{p}\cdot \boldsymbol{\alpha} + (m v^2 - B \; \boldsymbol{p}^2)\boldsymbol{\beta}
\end{eqnarray}

The equation reduces to the standard Dirac equation when $B=0$ and $v=c$, the speed of light. The spin is contained in the $\boldsymbol{\alpha}$ and $\boldsymbol{\beta}$ matrices and the parameter $B$ changes the effective mass of the $\boldsymbol{\beta}$ matrix term to make the $+m$ and $-m$ states dynamically different. The velocity $v$ indicates a many-body velocity associated with correlated behavior.  The 2-D modified Dirac equation reduces to a set of independent equations shown in Eq.\ref{topoDirac2D} (a 1/2 factor has been inserted for convenience in the definition of $B$).

\vspace{-0.0cm}
\begin{eqnarray}
\hspace{-0.4cm}\left[ v p_1 \sigma_x + v p_2 \sigma_y+\left(m v^2 -  \frac{B}{2} (p_1^2+p_2^2)\right) \sigma_z \right] \phi(x,y) = 0
\label{topoDirac2D}
\end{eqnarray}

In 2-D, there are two choices for describing the Hamiltonian. The first one is with $p_1=p_x$ and $p_2=p_y$, which leads to the standard 2-D edge state solution often published with a complex phase acting in the direction perpendicular to the edge state solution\cite{shen2012topological}. As a second choice, keeping the same spin orientation, we consider the same hamiltonian in a rotated reference frame from the first by setting $p_1=p_y$ and $p_2=-p_x$.  The sum of these two Hamiltonians, after reorganizing the terms in $p_x$ and $p_y$ ($\eta_1=\mp 1$), is a sum of two 1-D Hamiltonians shown in Eq.\ref{two1DDirac} (after a multiplication by $\sigma_x$). 

\vspace{-0.0cm}
\begin{eqnarray}
\left[ v p_1 \openone + \eta_1 i v p_1 \sigma_z- i \left(m v^2 -  B p_1^2\right) \sigma_y \right] \phi(x,y) = 0
\label{two1DDirac}
\end{eqnarray}

The sum of the two rotated Hamiltonians has achieved separation of variables i.e. $\phi(x,y)=\phi(x)\times \phi(y)$.  Setting $p_1=p_x$ with $\eta_1=-1$, and with a trial wave function $\phi(x)=C \chi_\eta \exp(-(\lambda_x + i  k_x) x)$ we obtain the modified secular polynomials shown in Eq.\ref{secularpolynomial1D} where we have isolated the complex part and, following Shen\cite{shen2012topological}, assumed the spinor in the wave function to represent a spin state being an eigenvector of $\sigma_y \chi_\eta = \eta \chi_\eta$ with $\eta=\pm 1$ (recall $\chi_\eta = (-\eta i,1)^T)$. We added $\eta_s=\pm 1$ for the action of $\sigma_z$ on this vector (it is diagonal and depends on the value of the spin eigenstate).

\vspace{-0.0cm}
\begin{multline}
\eta B \hbar^2 \lambda_x^2 - v \hbar\lambda_x + m v^2 - \eta_s  v \hbar k_x - \eta B \hbar^2 k_x^2 +...\\
...- i \left(v \hbar ( k_x - \eta_s \lambda_x) - 2 \eta  B \hbar^2\lambda_x k_x\right) = 0 
\label{secularpolynomial1D}
\end{multline}

Solving the real and imaginary part separately equal to zero leads to the two conditions for $\lambda_x$ and $k_x$ (with $\eta=sign(B)$ to keep the sign of $\lambda_x > 0$):

\vspace{-0.0cm}
\begin{equation}
\begin{split}
\lambda_x &= \frac{v}{2 \left|B\right| \hbar} \left[1 \pm \sqrt{1 - 4 m B +  \frac{ \eta_s 4 B \hbar k_x}{v} + \frac{4 B^2 \hbar^2 k_x^2}{v^2}}\right]\\
k_x &=  -\frac{ \eta_s \lambda_x  v}{2 \left|B\right| \hbar \lambda_x - v}
\end{split}
\label{secularpolynomial1Dfactors}
\end{equation}

These solutions are for the 1-D edge state solutions and are general for x and y (the y solution involves the change $\eta_s \rightarrow - \eta_s$). To obtain the bulk state solutions we set a closure of the gap $B \rightarrow 0$ and the two factors in Eq.\ref{secularpolynomial1Dfactors} reach the limit $\lambda_{x,+} \sim v/(\left|B\right| \hbar)$, for which $k_{x,+} \sim \eta_s v / ( \left|B\right| \hbar )$ ($\lambda_{x,+}$ becomes infinite and this wave function goes to zero), and for $\lambda_{x,-} = (\eta mv - \eta_s  \hbar k_{x,-})/\hbar$ and $k_{x,-}=\eta \eta_s  \lambda_{x,-}$ which in turn when combined give $k_{x,-}= \eta_s  \eta m v/(2\hbar)$ and  $\lambda_{x,-}=\eta mv/(2\hbar)$. We change the variable interpretation from Shen\cite{shen2012topological} by noting that $m v$ has the unit of momentum and define $Q= mv/\hbar$ {\color{black}which is the inverse of the topological correlation length}. The 1-D solution for $B \rightarrow 0$ of the superposed 2-D topological Dirac hamiltonians is given in Eq. \ref{topoDirac2Dwavefunction} where we have added phase difference $\phi$ to the different $k_x$ solutions (we integrate the $\phi(y)$ wave function assumed to be a bulk state with $m \rightarrow 0$).

\vspace{-0.0cm}
\begin{eqnarray} \label{topoDirac2Dwavefunction}
\hspace{-0.3cm}\phi(x)=C'  \left(\begin{matrix} \eta \; e^{ -i  \eta Q x/2}\\ i \;e^{ i  \eta Q x/2+i \phi}\end{matrix}\right) e^{- Q x/2}
\end{eqnarray}

The solution in the spinor has Cooper pairing character similar to the one seen in superconductivity (product of wave functions). Also, the magnetization from the spinor (subtraction of the wave functions) with each value of $\eta$ and $\phi=\pi/2$ reproduces a sinusoidal SDW type of magnetic ordering {\color{black}with a topological correlation length (decaying exponential) and undefined Q value} as shown in Eq. \ref{topoDirac2DwavefunctionSDW}.

\vspace{-0.0cm}
\begin{eqnarray} \label{topoDirac2DwavefunctionSDW}
\hspace{-0.3cm}\phi(x)= C \cos(Q x/2) e^{- Q x/2}
\end{eqnarray}

This is a remarkable result especially related to the second condition to achieve the bulk state, namely when $m \rightarrow 0$. First, if $m$ is not zero and if we take the $Q$ to represent a reciprocal lattice vector from a periodic crystal, the 1-D topological Dirac equation describes {\textit{antiferromagnetism}} per unit cell (because the reciprocal lattice vector in the phase factor modulating the spin state is divided by 2). This edge state antiferromagnetism has a magnetic form factor, the decaying exponential, also dependent on the same wave vector. On the other hand, if $Q$ is zero (which is achieved when $m \rightarrow 0$), the topological Dirac equation describes \textit{ferromagnetism} (all wave functions are added in phase at each unit cell) without decaying function hence achieving a bulk state for which there is no x-dependence in the wave function.  These two types of magnetic ordering are a direct consequence of the topological Dirac equation controlled by only one parameter, the many-body mass $m$, if $Q$ is chosen to be a reciprocal lattice wave vector. Ferromagnetism is also realized if $Q$ is an even multiple of a crystal lattice wave vector.

By subtracting the Eq. \ref{topoDirac2Dwavefunction} edge states solutions of different $B$ gap signs $\eta$, the spinor spin state can be expressed by a simple trigonometric function for which the net magnetization from the spinor is an exponential {\color{black}plane wave} function if an extra phase of $\phi=\pi/2$ exists between the two spin eigenstates as shown in Eq.\ref{topoDirac2Dcomplexwavefunction}.

\vspace{-0.0cm}
\begin{eqnarray} \label{topoDirac2Dcomplexwavefunction}
\hspace{-0.3cm}\phi(x) = C \; e^{ i Q x/2} \; e^{- Q x/2}
 \end{eqnarray}

We are interested in IC magnetic ordering which, as we show below, is obtained after a center of momentum transformation on some crystalline momenta $Q=\eta_m mv/\hbar$ and set $Q'=(Q - q_{cm})$, the crystal momenta expressed in the common center of momentum traveling at $q_{cm}$. The boson superposition of N $\phi(x)$  Eq.\ref{topoDirac2Dcomplexwavefunction} edge states involved in a many-body correlation is a product of wave functions in the center of momentum given by $\Psi_{cm}=\Psi_{cm,0} \;  e^{ i p'_t x}  \; e^{-Q'_t x}$ where $Q'_t=p'_t=\sum_{j} (\eta_{m,j} Q_j-q_{cm})/2$. The topological bulk state on $\Psi_{cm}$ requires that $m \rightarrow 0$ (to remove the x dependence of the decaying factor).  This condition is equivalent to $Q'_t \rightarrow 0$ which leads to $q_{cm}=(1/N)\sum_{j} \eta_{m,j} Q_j$ and sets both $Q'_t=p'_t=0$. The massless bulk state is ferromagnetic in the center of momentum which sets $q_{cm}$ as an average momentum. When expressed back in the crystal reference frame however, only the representation of the translation operator, the complex exponential with $p'_t=0$ in the center of momentum, will acquire a net momentum $p_t=0+q_{cm}$ in the crystal reference frame. The wave function per unit cell in the crystal reference frame becomes: $\Psi=\Psi_0 \; e^{ i q_{cm} x}$ (once established, the bulk state superposition with $Q'_t=0$ is invariant upon this transformation). This wave function has a finite lifetime because it is formed in a moving reference frame and is expected to fluctuate. 



As we show in the next section, the average momentum $q_{cm}$ corresponds to an IC magnetic ordering wave vector when the $Q_j$'s are taken to be local crystallographic planes or features in a crystal lattice {\color{black}which defines them as massive topological excitations}. 

It is interesting to note that the Eq.\ref{topoDirac2Dwavefunction} may have connection with superconductivity.  By summing the wave functions in the spinor we have generated different types of magnetic order and shown that IC magnetic order depends on a center of mass transformation. If we apply a direct product on the wave functions in the spinor of Eq.\ref{topoDirac2Dwavefunction}, the wave function can now describe superconductivity (because it pairs ($-\text{k} \uparrow \text{and k} \downarrow$) and ($-\text{k} \downarrow \text{and k} \uparrow$) depending on the $\eta$ solution chosen) but with a finite correlation length (decaying exponential). To have long range superconducting state and remove this decaying exponential, we need to apply a center of mass transformation on the pair. This is reminiscent of the Fulde-Ferrell-Larkin-Ovchinnikov (FFLO) superconducting state\cite{matsuda2007fulde} where the Cooper pair is ($-\text{k+q} \uparrow \text{and k} \downarrow$) in the crystal system but has zero net momentum in the center of mass.

\section{\label{sec:level1}Phenomenological model for IC magnetic order}

The calculation of the average wave vector $q_{cm}$ is phenomenological because there is no prior knowledge of which {\color{black}$Q$}-periodicities are at play in a crystalline lattice but it {\color{black}is} observed in several compounds representing different space groups. The summary of the following discussion appears in Table \ref{tab:table2}. 

\begin{table}
\vspace{-0.2cm}
\caption{\label{tab:table2}Summary of the phenomenological model for the calculation of the magnetic ordering wave vector $q_{cm}$ in several itinerant and metallic compounds compared with the experimental value.}
\begin{ruledtabular}
\begin{tabular}{lcccc}
  & Experimental & Calculated & \\
  & Wave vector & Wave vector & Error\\
\hline
Cr & 0.9515 & 0.95129 & 0.02\%\\
$\gamma$-Fe & 0.1 & 0.09722 & 3\%\\
CePtSn & 0.466&  0.46581  & 0.04\%\\
CePtSn (low T) & 0.418&  0.41978  &0.4\%\\
CePdSn  &0.473&  0.47280 &  0.04\%\\
CeRhIn$_5$ &  0.297&  0.29689  &0.04\%\\ 
CeRhSi$_3$ &  0.215& 0.21473  &0.1\%\\
CeNiAsO &  0.444&  4/9   & - \\
BSCCO &  0.21&  0.20833   & 0.8\% \\
MnSi &  0.024&  0.02325   & 3\% \\
\end{tabular}
\end{ruledtabular}
\end{table}

{\color{black}Cr has long been known to exhibit IC ordering and has been intensely studied by neutron\cite{fincher1979magnetic}$^,$\cite{koehler1966antiferromagnetism} and X-ray scattering experiments\cite{gibbs1988high}$^,$\cite{hill1995x}. In reciprocal space, a low temperature SDW is observed at Q=(0.9515,0,0) with $\tau$=0.0485 near the AF (1,0,0) zone center, a CDW at 2Q= (1.903,0,0) or (2-2$\tau$,0,0) is observed as well as another CDW harmonic at 4Q, which can be relabeled as (4-4$\tau$,0,0), also observed at 4Q-2=(2$\pm$4$\tau$,0,0) positions. A 3Q wave vector harmonic has also been observed\cite{Pynn3Qcr}. In the following analysis, simple crystallographic considerations can help gain insight about the experimental IC wave vectors observed. Cr has a BCC crystal structure Im$\bar{3}$m $\#$229 with atoms at $\left<0,0,0\right>$ and $\left<1/2, 1/2, 1/2\right>$ above T=312K and experience structural phase transitions below T=122K\cite{janner1980symmetry} to a tetragonal phase I4/mmm \#139 also with atoms at the (2a) symmetry positions $\left<0,0,0\right>$ and $\left<1/2, 1/2, 1/2\right>$ where the longitudinal incommensurate SDW develops. The tetragonal distortion does not change the reduced unit cell. For the phenomenological model, we use the positions of the atoms along the 1-D for 3 unit cells along the x-direction: x= 1/2, 1, 3/2, 2, 5/2, 3. The $Q$-vectors (the reciprocal 1/x) associated with these positions are: 2+0, 1+0, 2/3, 1/2, 2/5, 1/3. Taking the partial averages after n-terms of these Q-values we get: 2, 3/2, 1+2/9, 1+0.04167, 0.91333 (1-0.08667), 0.81667 (1-0.18333)} where we have isolated the fractional part from the reciprocal lattice wave vectors. We note that the last 3 averages of the list are a multiple of $\tau$: $\tau_1$=0.04167, $\tau_2$=0.08667/2=0.04333, $\tau_3$=0.18333/3=0.06111. Now taking the average of these three satellites in a two-stage averaging: ($\tau_1+\tau_2+\tau_3$)/3=0.04870 (1-0.95129), which is very close to the 0.9515 from the neutron scattering experiment\cite{fincher1979magnetic} (error of 0.02\%).

As another example, {\color{black}$\gamma$-Fe has a FCC crystal structure with a magnetic ordering wave vector of ($\tau$=0.1, 0, 1)\cite{PhysRevB.62.5564}.} The FCC unit cell has two positions along x = 0 and 1/2. First, let's consider the following sequence x= 0, 1/2, 1, 3/2. The cumulative partial average on the satellites of the associated $Q$-vectors 0, 2+0, 1+0 and 2/3 (0+1/3) leads 0, 0, 0, 1/12. On the other hand, if we take the sequence of atoms at x = 1/2, 1, 3/2 and the associated $Q$-vectors 2+0, 1+0 and 2/3 (0+1/3), the partial averages on the satellites are 0, 0 and 1/9. Now taking the two-stage averaging of the two sequences 1/12 and 1/9, we get: 7/72 = 0.09722 which is very close to the observed wave vector of $\tau$=0.1.

In the CePtSn, CePdSn series, the crystal structure is that of the space group Pna2$_1$ labelled \#33. Published work \cite{higashi1993crystal} chooses the space group Pn2$_1$a which is a permutation of the y- and z-axes. In this low-symmetry structure Pn2$_1$a has only one (4a) $\left<x,y,z\right>$ 
and there are only two positions along the y-axis: $y$ and $y+1/2$ per unit cell. This structure has no inversion symmetry and there are no atoms at $\left<0,0,0\right>$.

CePtSn has two magnetic phases\cite{kadowaki1994neutron}. The low temperature phase, below T$_M$, orders with a magnetic wave vector k=0.466b*. The Pt atoms along the b*-direction are at y$_{Pt}$=0.24275 and y$_{Pt}$+1/2. The reciprocal of these Pt crystallographic planes positions leads to 4.11946 and 1.34634 wave vectors. The average of the fractional part of these wave vectors is $\tau$=0.23290 which is half of the wave vector observed. Taking the $2^{nd}$-harmonic of the wave vectors we obtain 8.23893 and 2.69270 and since both wave vectors can be translated to zero because of the even integer parts, the predicted average satellite wave vector associated with the Pt atoms is 0.46581, very close to the experimental 0.466 (within 0.04\%) at low temperature.  {\color{black}The second harmonic of the crystal lattice wave vectors can be understood by putting 4Q (a ferromagnetic excitation with an even multiple of the crystalline wave vectors) in the Eq.\ref{topoDirac2Dwavefunction} wave function which leads to wave functions with a periodicity of 2Q and take these resulting wave vectors to build the average in a two-stage reference frame transformation.}

At high temperatures, between T$_M$ and T$_N$, CePtSn orders magnetically with a wave vector (h,k,l) with k=0.418b*. The Sn-atoms are at y$_{Sn}$=0.24627 and y$_{Sn}$+1/2. The $2^{nd}$-harmonics of the wave vectors from these positions are are 8.12117 and 2.68000. Translating the wave vectors in the same zone, from 0 to 1, the average on the fractional part is 0.40058 for the magnetic ordering wave vector compared to 0.418 observed experimentally. The difference of 4\% can be further improved to $\tau$=0.41978 (a difference of 0.4\%) taking the mixing between Pt and Sn sites as y$_{Pt}$, y$_{Sn}$, y$_{Pt}$+1/2, y$_{Sn}$+1/2, y$_{Pt}$+1, y$_{Sn}$+1 with the associated $2^{nd}$-harmonic wave vectors 8.23893, 8.12117, 2.6927, 2.68, 1.6093 (2+0.3907), 1.60479 (2+0.39521). These results are attributed to the mixing of the Pt to the Sn crystallographic planes at high temperature which vanishes at low temperature where the Pt crystallographic planes alone drive the magnetic phase in CePtSn. It is worth noting that the $\tau$ for both magnetic phases require the $2^{nd}$-harmonics of the wave vectors and for an expansion of the satellites near an 'even' reciprocal lattice wave vector (with the same chirality, i.e. the positive satellites).

The case of CePdSn involves more atoms without taking the crystal wave vectors harmonics. Neutron scattering experiments\cite{kohgi1992neutron}$^,$\cite{kasaya1991crossover} have shown the magnetic wave vector is k=0.473b* below $T_N=7K$. Taking the Pd atoms, y$_{Pd}$=0.24414, y$_{Pd}$+1/2, y$_{Pd}$+1 and y$_{Pd}$+3/2 leads to the wave vectors 4.09601, 1.34383 (2+0.65617), 0.80377 and 0.57335. Doing the same with the Sn sites, y$_{Sn}$=0.24674, y$_{Sn}$+1/2, y$_{Sn}$+1, y$_{Sn}$+3/2, the wave vectors are 4.05285, 1.33915 (2+0.66085), 0.80209 and 0.57250 respectively. The average over the eight Pd and Sn atoms is 0.52720 = 1-0.47280 (to respect the published labelling) which is now very close to the experimental value  (within 0.04\%) underlying the role of the Pt and Sn crystallographic planes averaging in this series of compounds.

Focusing on other orthorhombic compounds, CeRhIn$_5$ is a heavy fermion with a tetragonal P4/mmm \#123 crystal structure with Ce atoms at the (1a) $\left<0,0,0\right>$ symmetry position, the Rh at the (1b) $\left<0,0,1/2\right>$ symmetry position, and the In atoms at the (1c) $\left<1/2,1/2,0\right>$ and at the (4i) $\left<0,1/2,z_{In}\right>$, $\left<1/2,0,z_{In}\right>$, $\left<0,1/2,-z_{In}\right>$, $\left<1/2,0,-z_{In}\right>$ positions with $z_{In}$=0.30592 as published\cite{moshopoulou2002comparison}. It orders magnetically\cite{bao2000incommensurate} below T$_N$=3.8K with an IC wave vector (1/2, 1/2, 0.297). With the $z_{In}$ components of the (4i) sites, there are two relevant positions:  $z_{In}$=0.30592 and 1-$z_{In}$=0.69408 in the folded unit cell. The $2^{nd}$-harmonic IC wave vector from these positions are: 6.53766 and 2.88151. The average satellite is 0.709595 (1+0.290415) which compares favorably to the 0.297 ordering wave vector, an error of 2\%. In fact, the sum converges to 0.29689 (an error of 0.04\%) after 7 crystallographic planes if we consider the positions: $z_{In}$, 1-$z_{In}$, 1+$z_{In}$, 2-$z_{In}$, 2+$z_{In}$, 3-$z_{In}$, 3+$z_{In}$ with the $2^{nd}$-harmonic wave vectors: 6+0.53766, 2+0.88151, 2+0.46513, 2+0.819418, 0.867333, 0.742368, 0.604975.

Another tetragonal heavy-fermion compound CeRhSi$_3$ has a noncentrosymmetric I/4mm \#107 crystal structure\cite{kimura2010novel}. Its magnetic ordering\cite{ASO2007602} is IC with a wave vector of $(\pm \delta, 0, 1/2)$ with $\delta=0.215$ below T$_N$=1.6K\cite{doi:10.1002/pssb.201200772}. The atoms are either at the (4b) $\left<0,1/2,z\right>$ and $\left<1/2,0,z\right>$ or at the (2a) $\left<0,0,z\right>$ which leaves only atoms at x=0 or 1/2 along the x-direction. The IC wave vector of 0.21473, within 0.1\% of the experimental 0.215, is found taking the 1/2-harmonic (AF structure) of the 8 crystallographic planes at $x=1/2$, $1$, $1/2+1$, $1+1$, $1/2+2$, $1+2$, $1/2+3$, $1+3$ which leads to the same chirality wave vectors 1+0, 1/2, 1/3, 1/4, 1/5, 1/6, 1/7 and 1/8. Note that we added a '0' in the averaging which stands for electrons matching the periodicity of the lattice. {\color{black}The 1/2-harmonic could be explained by the development of Eq.\ref{topoDirac2Dwavefunction} wave functions with the lattice wave vectors Q first, which leads to a periodicity of Q/2, and then from these wave wave vectors build the average $q_{cm}$ in a two stage reference frame transformation.}

The phenomenological model gives some insight into the IC order in the heavy fermion CeNiAsO. It has a P4/nmm \#129 (origin \#2) crystal structure with the Ce atoms at the Wyckoff (2c) positions $\left<1/4, 1/4, z_{Ce}\right>$ with z$_{Ce}$=0.1465, the Ni atoms at the (2b) positions $\left<3/4,1/4,1/2\right>$, the As at the (2c) positions $\left<1/4, 1/4, z_{As}\right>$ with z$_{As}$=0.6434 and the O atoms at the (2a) positions $\left<3/4, 1/4, 0\right>$ as published\cite{0953-8984-23-17-175701}. CeNiAsO experiences an IC phase transition attributed to a SDW below T=7.6K with a wave vector (0.444, 0, 0)\cite{wu2017incommensurate}. The atomic positions involve a redundancy of 1/4 and 3/4 coordinates along the $\vec{a}$-direction shared by the metallic ions: the Ce atoms position at 1/4, 3/4 and the Ni positions at 3/4, 1/4. Taking 1/4, 3/4, 3/4 for the unit cell leads to the reciprocal wave vectors 4+0, 4/3 (2+2/3), and 4/3 (2+2/3). Taking the average over the three satellites (0 + 2/3 + 2/3)/3=4/9 which is exactly the reciprocal wave vector found experimentally, highlighting a mixing between the metallic character of the Ce and Ni atoms. It is worth mentioning that CeNiAsO has no superconductivity detected down to 30mK\cite{0953-8984-23-17-175701} and that topological effects may act as to prevent such occurrence due to added insulating fluctuations.

A similar analysis can be made in the high-Tc superconductor Ba-Sr-Ca-Cu-O 2212.  It exhibits an IC wave vector\cite{gao1988incommensurate} at $\tau$=$\pm$0.21a*. BSCCO has a Amaa \#66 space group crystal structure with atoms along the $\vec{a}$-direction: x, -x, 1/2+x, 1/2+x, -x, x, 1/2-x, 1/2-x and the same atoms translated by (0, 1/2, 1/2). Here all the atoms in BSSCO are at either x=0, 0.25, 0.5, 0.75 positions. The metallic Cu atoms are at the general position $\left<0.5, 0.2498, 0.1967\right>$ with $x=0.5$. Taking the non-equivalent position (except the zeroes): 0.5, 1, we get the expanded unit cell 0.5, 1, 0.5+1=3/2, 1+1=2 and obtain the wave vectors 2+0, 1+0, 2/3 (0+1/3) and 0.5 (0+1/2) which leads to the average over the 4 satellites (0+0+1/3+1/2)/4=0.20833, very close to 0.21 published (an error of 0.8\%).

MnSi has attracted a lot of attention recently because it can be considered a topological insulator\cite{neubauer2009topological} and exhibits a Skyrmion lattice phase upon application of a magnetic field\cite{day2009exotic}. Its crystal structure\cite{nakanishi1980origin} belongs to the tetrahedral $P2_13$ \#198 space group with both Mn and Si ions occupying the (4a) symmetry position $pos_1=\left<u,u,u\right>$, and the three coplanar sites perpendicular to the $\left<1,1,1\right>$ direction $pos_2=\left<-u+1/2,-u,u+1/2\right>$, $\left<-u,u+1/2,-u+1/2\right>$,$\left<u+1/2,-u+1/2,-u\right>$ with $u_{Mn}=0.138$ and $u_{Si}=0.845$.  Below $T_N$=29.5K, the onset of a helical SDW is observed\cite{muhlbauer2009skyrmion} with a wavelength of 190\AA  \; along the $\left<1,1,1\right>$ direction which can be viewed, in reduced reciprocal lattice units, as $2\pi/190 = h \; 2\pi/a$ with $a=4.56\AA$  \; and $h=0.024$. The reciprocal space length $\sqrt{3}/\left(pos_j \cdot \left<1,1,1\right>\right)$ of the (4a) lattice sites along diagonal direction leads to two wave vectors: $1/(\sqrt{3}u)$ and $\sqrt{3}/(1-u)$, the latter coming from the three coplanar sites. The lack of inversion symmetry of the $\left<u,u,u\right>$ sites leads to the wave vectors $1/(\sqrt{3}(u+n))$ with $n$ an integer sites after a $\left<n,n,n\right>$ translation along the diagonal. Taking $n$=-4,-3,-2,-1,0,1,2,3 (a symmetric set), we obtain the wave vectors (keeping different  $\eta_m$ signs for the $Q_j$):  -0.14950, -0.20173, -0.31007, -0.66978, 4+0.18370, 0.50734, 0.27004, 0.18399 for an average of $q_{cm}=-0.02325$ which is very close to the published $h=0.024$ satellite (an error of 3\%) for local correlations in the helical phase.


\section{\label{conclustion}Conclusion}

The 1-D topological edge state solutions per unit cell from two rotated 2-D Dirac Hamiltonians on a discrete lattice offer a general framework to describe the magnetic order in itinerant/metallic systems depending on one parameter, the mass $m$. In a broader sense, these solutions qualitatively describe the evolution of ferromagnetic order in metallic compounds\cite{brando2016metallic} upon cooling temperatures when it is interrupted at a tricritical point to transform into an antiferromagnetic/IC magnetic order as the competition between the creation of topological bulk states and massive excitations appear due to the presence of a lattice. The phenomenological model built from the edge state solutions reproduces the IC magnetic wave vectors seen in diffraction experiments as an average of wave vectors based on the {\color{black}lengths scales provided by a} crystalline lattice. Due to their crystal lattice origin, these IC wave vectors are not expected to move with temperature, except when there is a change in the number of electrons involved in the average, which may change with temperature. The algorithm to calculate the IC wave vector is reduced to simple reciprocal space concepts like: a local expansion for the wave vectors using crystallographic planes as a basis, translations in reciprocal space for the satellites with the same chirality and taking a harmonic of the crystal lattice wave vectors. A Python script casting these concepts is available by contacting the author. I acknowledge the support from Kent State University.
\vspace{-0.0cm}

%

%

\end{document}